\documentclass{article}
\usepackage{ijcai24}

\usepackage{times}
\usepackage{soul}
\usepackage{url}
\usepackage[hidelinks]{hyperref}
\usepackage[utf8]{inputenc}
\usepackage[small]{caption}
\usepackage{graphicx}
\usepackage{amsmath}
\usepackage{amsthm}
\usepackage{booktabs}
\usepackage{algorithm}
\usepackage{algorithmic}
\usepackage[switch]{lineno}
\usepackage{xcolor}

\title{MR-MT3: Memory Retaining Multi-Track Music Transcription to Mitigate Instrument Leakage}

\author{
Hao Hao Tan$^{1*}$
\and
Kin Wai Cheuk$^{2*}$\and
Taemin Cho$^1$\and
Wei-Hsiang Liao$^{2}$\and
Yuki Mitsufuji$^{2,3}$\\
\affiliations
$^1$BandLab, Singapore\\
$^2$Sony AI, Japan\\
$^3$Sony Group Corporation, Japan\\
\emails
haohao.tan@bandlab.com,
kinwai.cheuk@sony.com
}

\begin{document}

\maketitle

\begin{abstract}

This paper presents enhancements to the MT3 model, a state-of-the-art (SOTA) token-based multi-instrument automatic music transcription (AMT) model. Despite SOTA performance, MT3 has the issue of instrument leakage, where transcriptions are fragmented across different instruments. To mitigate this, we propose MR-MT3, with enhancements including a memory retention mechanism, prior token sampling, and token shuffling are proposed. These methods are evaluated on the Slakh2100 dataset, demonstrating improved onset F1 scores and reduced instrument leakage. In addition to the conventional multi-instrument transcription F1 score, new metrics such as the instrument leakage ratio and the instrument detection F1 score are introduced for a more comprehensive assessment of transcription quality. The study also explores the issue of domain overfitting by evaluating MT3 on single-instrument monophonic datasets such as ComMU and NSynth. The findings, along with the source code, are shared to facilitate future work aimed at refining token-based multi-instrument AMT models.
\end{abstract}

\section{Introduction}\label{sec:intro}
\def\thefootnote{*}\footnotetext{Equal contribution}\def\thefootnote{\arabic{footnote}}
Automatic music transcription, also known as AMT~\cite{benetos2018automatic}, is a process of converting the audio signal of a music performance into a symbolic representation, typically in the form of sheet music or MIDI data. The primary goal is to accurately capture the musical notes, timing, and other expressive elements of a music piece. This task comes with challenges such as polyphony, where multiple notes can be played simultaneously. Moreover, precise time annotation is required to train a model to detect the extract onset location for pitches. Automatic piano transcription~\cite{hawthorne2017onsets,kelz2019deep,pedersoli2020improving,cheuk2021reconvat,kong2021high,cheuk2023diffroll} is one of the mostly studied case for single instrument polyphonic music.

However, most music audio contains more than one instrument. This requires simultaneous transcription of multiple instruments playing together in an ensemble or mix. The objective is to extract the individual parts for each instrument from the combined audio signal. Challenges in this context include instrument separation, dealing with the timbral complexity of different instruments, and addressing variations in the arrangement of instruments. The applications of this field range from enhancing music production~\cite{hawthorne2022multi} through enabling MIDI-level rearrangement for audio-only mixes, to providing educational tools for musicians to learn specific parts of a composition. It also plays a crucial role in archiving~\cite{Wongsaroj} and analyzing musical recordings~\cite{Ewert,ru2021computer} for research and archival purposes. 

Currently, only a handful of models are capable of tackling multi-instrument music transcription~\cite{tanaka2020multi,wu2020multi,gardner2021mt3,lu2023multitrack,cheuk2023jointist}. Among all of them, MT3 ~\cite{gardner2021mt3} is special in a way that it models AMT as a language task by predicting music event tokens in an autoregressive manner given the input spectrogram. 



\begin{figure}[t]
\centering
\includegraphics[width=\columnwidth]{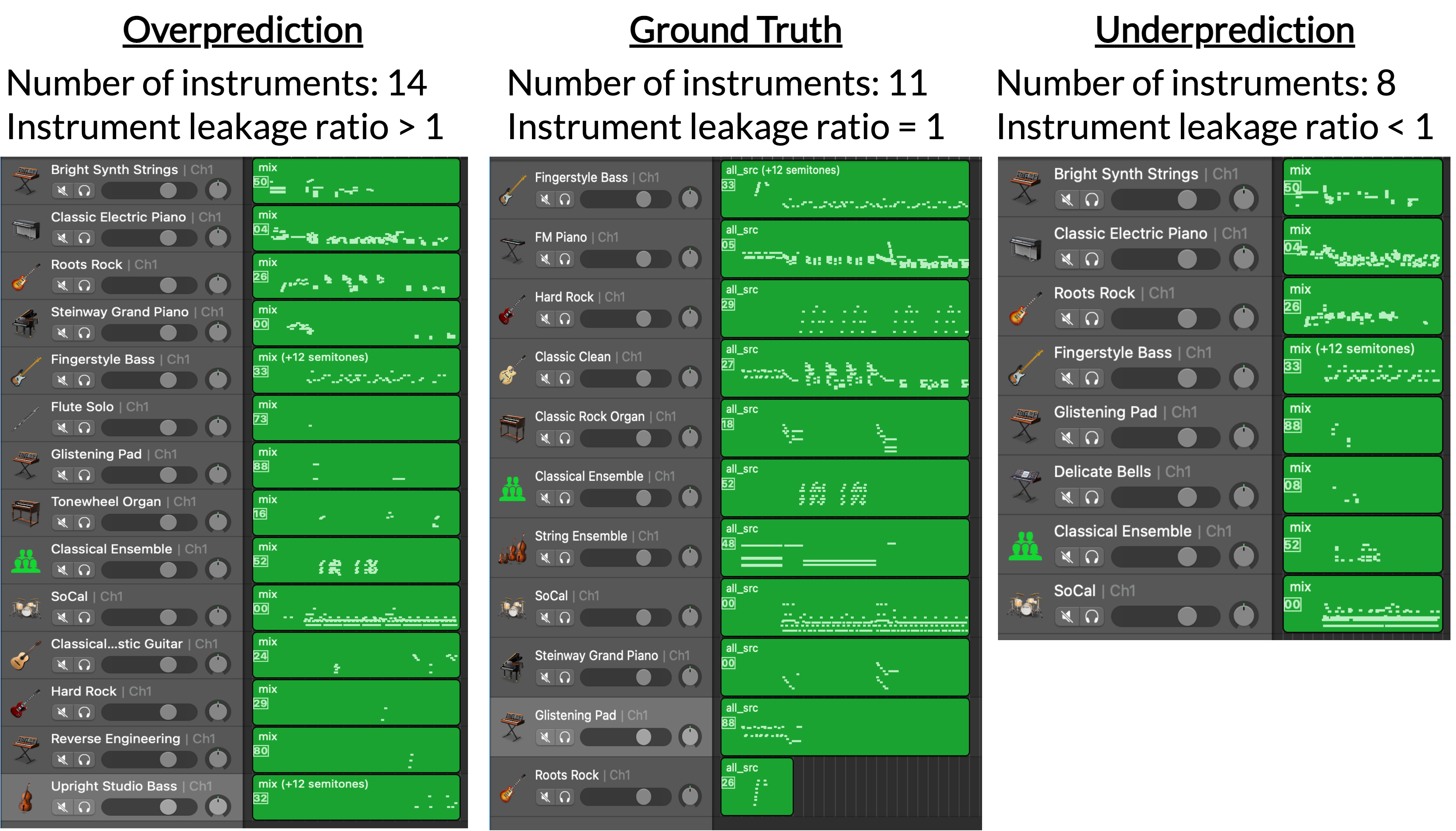}
\caption{An example of the instrument leakage issue. In the MIDI transcribed by MT3 (left), we often observe musical notes intended for a specific instrument to ``leak'' across multiple instruments, leading to a cluttered arrangement as compared to the ground truth (middle). We also demonstrate a transcription example which under-predicts the number of instruments (right).}
\vspace{-5mm}
\label{fig:instrument_leakage}
\end{figure}

\begin{figure*}[t]
\centering
\includegraphics[width=2\columnwidth]{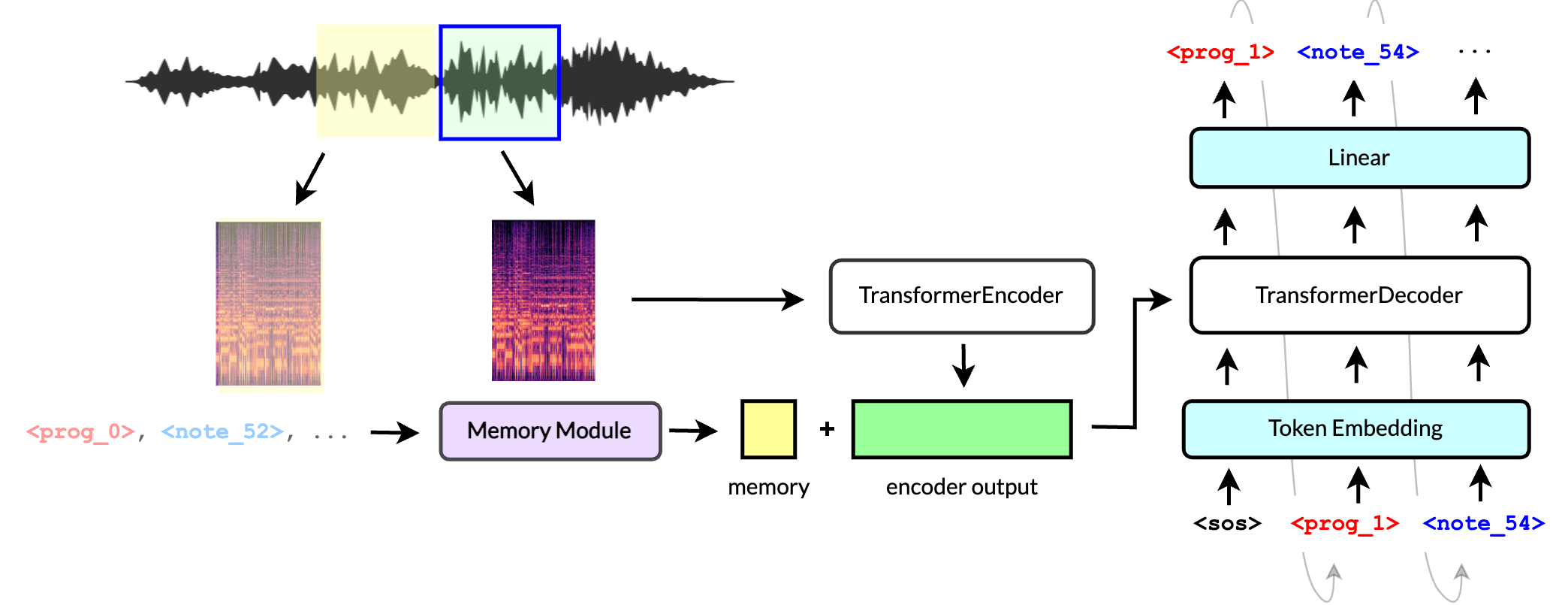}
\vspace*{-3mm}
\caption{Our proposed model architecture for MR-MT3. A memory retention mechanism is introduced to aggregate tokens transcribed from the previous segment (yellow). It is concatenated to the encoder outputs for cross-attention during autoregressive token sampling of the current segment (green).}
\label{fig:model_arch}
\end{figure*}

In this paper, we aim to enhance MT3, which, despite achieving state-of-the-art transcription accuracy, has limitations. Specifically, MT3 splits audio into non-overlapping segments, and transcribes each segment independently, thereby potentially loses the inter-dependencies between segments. Music, however, possesses a long-term structure. The inability to maintain the musical context shared across audio segments may result in a transcription that lacks melodic continuity and coherence in harmony and instrumentation. One such incoherence issue we observe in MT3 transcription is depicted in the left side of Figure~\ref{fig:instrument_leakage}, whereby the melody is played by one instrument across several audio segments, but its transcription is fragmented across different instruments in each segment. We refer to this phenomenon as \textbf{instrument leakage} in this study. To address this, we propose MR-MT3 (Memory Retaining MT3), by applying the following ideas:

\begin{itemize}
  \item \textbf{Memory retention}: the aggregated representation from the past segment is used to inform the transcription of the current segment using a memory module.
  \item \textbf{Prior token sampling}: during training, the past segment used for memory retention can be sampled from prior frames up to a certain time horizon, instead of only taking the immediate past segment.
  \item \textbf{Token shuffling}: semantic-preserving token shuffling methods as a data augmentation method for training.
\end{itemize}

As of this writing, token-based automatic music transcription remains a relatively innovative concept. We anticipate that our findings and results will provide valuable insights for future research in this area.


\section{Related Work}\label{sec:related}

Automatic music transcription (AMT) is traditionally modeled as the task of converting a given spectrogram into its corresponding piano roll. Usually, the piano roll has the same time resolution as the spectrogram, making it a task that resembles image segmentation~\cite{wu2019polyphonic}. We denote this type of AMT systems as \textbf{piano roll-based} AMT. Using this definition, MT3 belongs to \textbf{token-based} AMT. The majority of the existing AMT systems are piano roll-based. Moreover, depending on the instrument granularity of the transcription, we can also classify AMT systems into \textbf{instrument-agnostic} and \textbf{multi-instrument} AMT. In the following subsection, we will introduce several existing works which depicts the above-mentioned types of AMT systems.


\subsection{Piano Roll-based AMT}
\subsubsection{Instrument-Agnostic}
Instrument-agnostic AMT models take in multi-instrument audio as input but output only one channel of piano roll without the instrument information. These models transcribe pitches from the audio and disregards which instrument is playing the pitch. Examples include \textbf{MusicNet}~\cite{thickstun2016learning}, \textbf{DeepSalience}~\cite{bittner2017deep}, \textbf{BasicPitch}~\cite{bittner2022lightweight}, and \textbf{Timbre-Trap}~\cite{cwitkowitz2023timbre}.

\subsubsection{Multi-Instrument}
When the instrument information of the pitches is required, multi-instrument AMT is needed. Multi-instrument satisfies this requirement by outputting the number of channels of piano rolls that equal to the number of instruments. Then, the model separates the pitches corresponding to an instrument to its specific channel~\cite{wu2020multi,tanaka2020multi,lu2023multitrack}. However, these systems do not scale up well with the number of instruments, as the number of output channels also needs to be scaled up alongside it. \textbf{Jointist}~\cite{cheuk2023jointist} is designed to tackle this issue by taking in arbitrary number of instrument conditions, and transcribes the pitches into a single-channel piano roll for each given instrument condition.

\subsection{Token-based AMT}

MT3~\cite{gardner2021mt3} is a multi-instrument AMT model which distinguishes itself from other piano roll-based AMT systems by approaching music transcription as a language modelling task~\cite{chen2021pix2seq}. Rather than utilizing piano rolls for transcription, MT3 uses streams of event tokens, such as time token, program token, note-on/off token, and note token (illustrated in Figure~\ref{fig:token_shuffling}), to denote the musical events present in the audio. MT3 uses an encoder-decoder Transformer model, and uses spectrogram frames as input for the encoder. To decode the stream of events, the Start of Sequence (SOS) token is then provided to the decoder to predict the subsequent token in an autoregressive manner, based on the encoder output. As input audio is split into segments, MT3 also introduces a “tie” section, separated by a "tie" token, to declare notes that are held over from the previous segments. We refer the readers to section 3.2 of the MT3 paper~\cite{gardner2021mt3} for a detailed introduction of the proposed vocabulary.

This token-based methodology has demonstrated its effectiveness in piano transcription~\cite{Hawthorne2021SequencetoSequencePT}, and it also performs reasonably well in multi-instrument transcription. However, as mentioned in previous sections, it suffers from the instrument leakage problem. In the next section, we will review the technique developed in the field of computer vision to handle long term dependencies in videos. Based on this technique, we develop our own version for token-based AMT in section~\ref{sec:memory_module_amt}.

\begin{figure}[t]
\centering
\includegraphics[height=80mm]{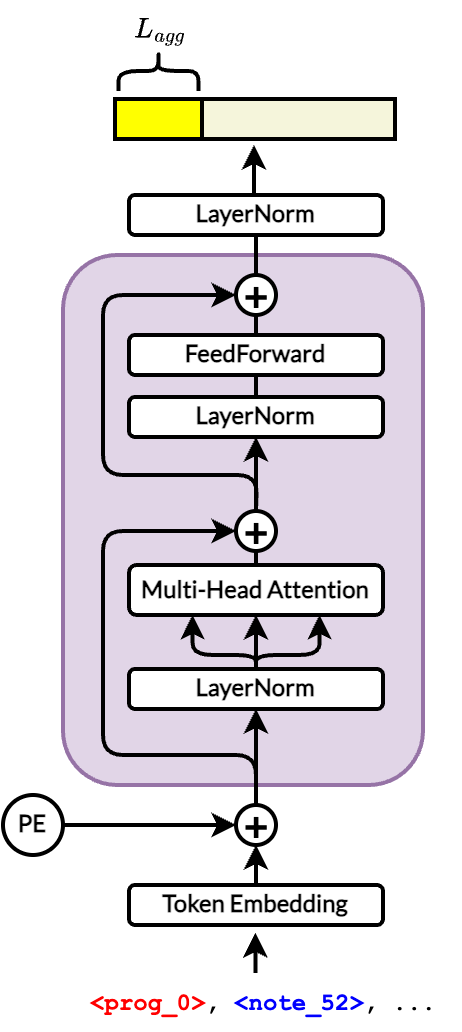}
\caption{The memory retention mechanism. The aggregated token representation is its self-attention output, truncated at length $L_\text{agg}$.}
\label{fig:memory_module}
\end{figure}

\subsection{Memory Retention Mechanism for Transformers}\label{sec:memory_module_transformer}
The memory retention mechanism is commonly used in the video processing domain to capture long term dynamics. For example, the Memory-Augmented Multiscale Vision Transformer (MeMViT)~\cite{wu2022memvit} proposes to treat a long video as a sequence of short clips, and process each clip sequentially. Previous clips are cached and compressed, so that the model processing the current clip can reference prior context at only a marginal cost. With better temporal support, MeMViT consistently achieves large gains in recognition accuracy. In video captioning, Memory-Augmented Recurrent Transformer (MART) ~\cite{lei2020mart} employs a memory module to generate a condensed memory state from the video segments and the sentence history. This facilitate better prediction of the subsequent sentence, thus encouraging coherent paragraph generation. Analogous to the task of video captioning, token-based AMT can also be considered as audio captioning, and the musical event tokens are analogous to the captions. Drawing on this analogy, we propose our own variant of a memory retention mechanism for token-based AMT in the subsequent section.


\section{Proposed Method}\label{sec:proposed}
In this section, we will elucidate each of the proposed ideas designed to tackle the instrument leakage issue.

\begin{figure}[t]
\centering
\includegraphics[width=\columnwidth]{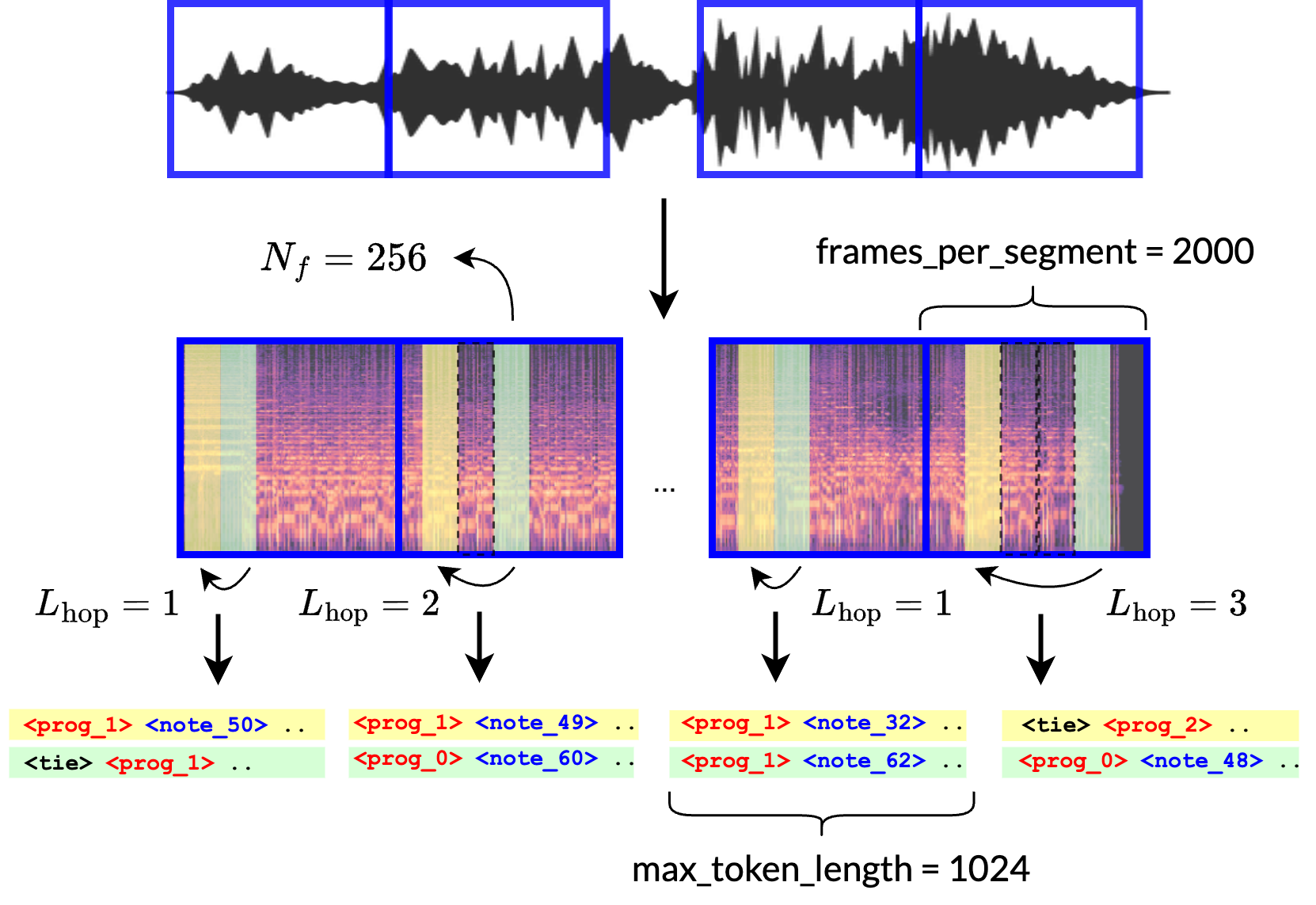}
\caption{Segmentation workflow to obtain training pairs following MT3. In addition, we propose to use the prior frames (yellow) to inform the transcription of the current frames (green). Prior frames can start from up to $L_\text{max\_hop} \times N_{f}$ before the current frames. In the above example, $L_\text{max\_hop} = 3$.}
\label{fig:prior_sampling}
\end{figure}

\begin{figure}[t]
\centering
\includegraphics[width=\columnwidth]{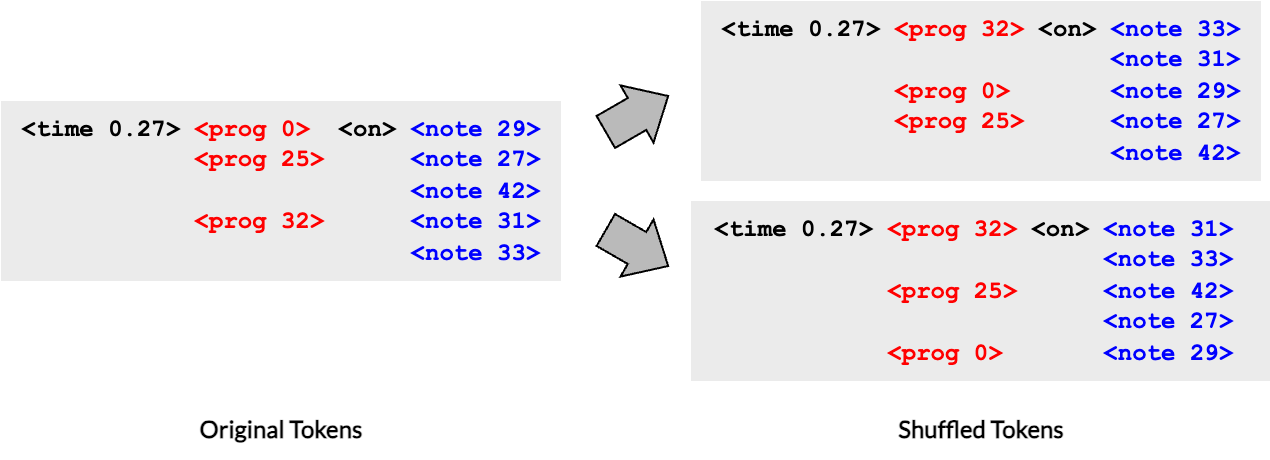}
\caption{Event token shuffling as a data augmentation method.}
\label{fig:token_shuffling}
\end{figure}

\begin{table*}[ht]
    \centering
    \begin{tabular}{lcccccc}
    \toprule
    & \multicolumn{3}{c}{Onset-Offset F1 Scores (\%)} & \multicolumn{3}{c}{Onset F1 Scores (\%)} \\
    \cmidrule(lr){2-4} \cmidrule(lr){5-7}
     & Flat & MIDI Class & Full & Flat & MIDI Class & Full \\
    \hline
    Reported       & 48   & \textbf{62}   & \textbf{55}   & \textbf{76.0} & N.A. & N.A \\
    Reproduced (c) & \textbf{50.9} & 48.9 & 25.8 & 76.0 & \textbf{73.0} & \textbf{38.9}   \\
    Reproduced (s) & 37.9 & 35.3 & 21.2 & 64.9 & 59.5 & 33.5  \\
    
    \hline
    \end{tabular}
    \caption{Onset and offset multi-instrument F1 scores (left) and onset multi-instrument F1 scores (right) reported in the MT3 paper compared to the reproduced number when training the model from scratch (s) and from the officially provided checkpoint (c).}
    \label{tab:reproduce}
\end{table*}
\subsection{Memory Retention Mechanism for AMT}\label{sec:memory_module_amt}

By incorporating prior knowledge of the past musical events, we hypothesize that it helps eliminating unlikely upcoming events and reducing instrument leakage. The memory retention mechanism takes in past $N_{t} = 1024$ event tokens and extract the embedding for each token via a trainable embedding layer as depicted in Figure~\ref{fig:memory_module}. The embedding is processed through a single layer of multi-head attention with 6 heads. Ultimately, $L_\text{agg}$ of the self-attention output is utilized as the memory block and it is concatenated with the Transformer encoder output. The memory length $L_\text{agg}$ is a hyper-parameter to be explored. 

During training, the memory retention mechanism is trained via teacher forcing. i.e., the ground truth labels of past tokens are used to extract the memory block of length $L_\text{agg}$. More details about prior token sampling will be discussed in the next subsection. During inference, we use the tokens predicted in the previous segment by the model as the prior tokens.

\subsection{Prior Token Sampling}
The prior token sampling method is illustrated in Figure~\ref{fig:prior_sampling}. Let $F[i:i+N_{f}]$ be the set of spectrogram frames used for the current training step (green region), starting from index $i$ and $N_f = 256$. Each group of frames correspond to $N_{t} = 1024$ event tokens. To sample prior tokens as memory, the most intuitive approach is to obtain $N_t$ tokens corresponding to the set of frames right before the current set, i.e. 
\begin{equation}
F[i-N_{f}:i]\,,
\end{equation}%
which is shown as the yellow region in Figure~\ref{fig:prior_sampling}. However, we argue that as long as the prior tokens are not too far away in the past, their context is still useful for transcribing the current frame. Therefore, we allow the sampling of prior tokens corresponding to the spectrogram frame $L_\text{hop}\times N_f$ before $i$, i.e.
\begin{equation}
F[i-L_\text{hop} \times N_{f} :i+(1-L_\text{hop})\times N_{f}]\,,
\end{equation}%
 During training, we introduce a hyper-parameter $L_\text{max\_hop}$, and the value of $L_\text{hop}$ is randomly sampled between $[1, L_\text{max\_hop}]$ when generating each training sample pair. We illustrate the effect of $L_\text{max\_hop}$ on the transcription results in Table~\ref{tab:result_scratch}.


\subsection{Token Shuffling}
We propose token shuffling as a data augmentation method to be applied to the tokenization mechanism proposed in MT3. In the original work, at a given timestamp, the tokenization follows the grammar of ascending order of program numbers and note numbers. However, within the same timestamp and the same note-on / note-off section, the program-note groups can be shuffled while preserving the same transcription. An example of shuffled tokens is illustrated in Figure~\ref{fig:token_shuffling}.

\begin{table*}[htbp]
  \centering
  \begin{tabular}{ccccccccccc}
    \toprule
    & & & & \multicolumn{3}{c}{Onset F1 Scores (\%)} & & \multicolumn{3}{c}{Instrument Detection (\%)}\\
    \cmidrule{5-7} \cmidrule{9-11}
    Model & Token Ordering & ${L_\text{agg}}$ & ${L_\text{max\_hop}}$  & \text{Flat} & \text{MIDI Class} & \text{Full} & \text{$\phi$} & \text{Precision} & \text{Recall} & \text{F1} \\
    \hline
    MT3 & Fixed & N.A & N.A & {63.3} & {57.9} & {32.7} & 1.70 & 27.8 & \textbf{46.0} & 34.0\\
    
    MT3 & Shuffled & N.A & N.A & 64.9 & 59.5 & 33.5 & 1.65 & 28.5 & 45.2 & 34.2\\
    MR-MT3 & Shuffled & 32 & 1 & 66.4 & 61.6 & 34.5 & \textbf{1.05} & \textbf{39.4} & 40.5 & \textbf{39.5} \\
    MR-MR3 & Shuffled & 64 & 1 & 66.9 & 62.0 & 35.0 & 1.12 & 37.6 & 41.6 & 39.0 \\
    MR-MR3 & Shuffled & 64 & 2 & 66.9 & 62.1 & 34.8 & 1.16 & 37.6 & 42.7 & 39.4 \\
    MR-MR3 & Shuffled & 64 & 3 & \textbf{67.3} & \textbf{62.5} & \textbf{35.3} & 1.24 & 36.3 & 43.8 & 39.1 \\
    MR-MR3 & Shuffled & 64 & 8 & 67.1 & 62.4 & \textbf{35.3} & 1.35 & 34.4 & 44.4 & 38.1 \\
    \bottomrule
  \end{tabular}
  \caption{Result for training the model from scratch}
  \label{tab:result_scratch}
\end{table*}
\section{Experiments}\label{sec:experiments}
\subsection{Dataset and Training}\label{sec:dataset}
We train our models on the redux version train set of Slakh2100 in which the duplicated tracks are removed from the train set, resulting in 1289 tracks in the train set instead of the original 1500 tracks. We keep our training setting mostly similar to MT3, as shown in Figure~\ref{fig:prior_sampling}: the audio clips are down-sampled to 16kHz and divided into $N$ segments of 256k audio samples. The audio segments are converted into log-scaled mel-spectrograms using an FFT length of 2048 samples, a hop size of 128, and 512  Mel bins. Hence, each audio segment corresponds to 2000 spectrogram frames. For each audio segment, $N_f = 256$ consecutive spectrogram frames are sampled, and the event tokens with a maximum length of $N_t = 1024$ are extracted to form a training sample pair. With the memory retention mechanism described in section~\ref{sec:memory_module_amt}, we carry over the labels corresponding to the prior segments and concatenate them with the encoder output. Due to memory constraints of the GPU, we randomly select 12 consecutive audio segments from a sample to form a batch for each training step. The model is trained for 800 epochs on two RTX 4090s, which takes approximately 2 days. We monitor the training using the validation set and save both the best checkpoint and the last checkpoint. Then we compare and check the F1 score of the checkpoints on the validation set. We find that the checkpoint with the lowest validation loss does not necessarily result in the best F1 score on the validation set. Most of the time, the last checkpoint has a better F1 score than the checkpoint with the lowest validation loss. Hence, we report the results on the test set using the last checkpoint for all of our experiments.



\subsection{Reproducing Baseline Results}\label{sec:reproduce}
We intend to improve upon MT3, however we face difficulties reproducing the same F1 scores reported in the MT3 paper~\cite{gardner2021mt3}. First, the official repository\footnote{https://github.com/magenta/mt3} does not provide the source code for model training, only the source code for inference and a pre-trained checkpoint are provided. We evaluate the multi-instrument onset-offset F1 scores on the pre-trained checkpoint, but we obtain lower scores than the reported numbers (the second row v.s. the first row of Table~\ref{tab:reproduce}). For example, the reported onset-offset F1 score for the MIDI Class instrument granularity is 62, while the released checkpoint yields only 48.9. The same goes for the Full instrument granularity, the released checkpoint yields only 25.8, much lower than the reported score, 55. We switch to evaluate the onset-only F1 score, and we obtain the same value as reported, 76, for the Flat granularity.

As training is not supported in the official repository, we develop our training code based on an open-source implementation\footnote{https://github.com/kunato/mt3-pytorch}. When we train the model from scratch using the same training hyper-parameter (a fixed learning rate of $1\times10^{-3}$) described in the MT3 paper, we get a much worse result. After doing a brief hyper-parameter search, we find that a learning rate of $2\times10^{-4}$ with a cosine scheduler of minimum learning rate of $2\times10^{-5}$ and 64500 warm-up steps yields better F1 scores (third row of Table~\ref{tab:reproduce}). However, we are still unable to obtain results close to the reported numbers.

Due to the above difficulties, we divide our experiments into two parts: training \textbf{from scratch}, and \textbf{continual training} from the provided checkpoint. Our goal is to apply our proposed methods to both scenarios to see if it improves the multi-instrument transcription results.

\textbf{From scratch:} We train the original MT3 from scratch for 800 epochs using a learning rate of $2\times10^{-4}$ with a cosine scheduler as mentioned above. This model is used as the baseline. Then we apply our proposed methods and benchmark the improvements against the baseline. 

\textbf{Continual training}: To provide a more comprehensive study on our proposed ideas, we initialize MT3 with weights derived from the checkpoint provided by the official repository\footnotemark[1]. We further train the model, with and without our proposed methods, for 100 epochs using a fixed learning rate of $1\times10^{-5}$. We briefly explore other hyper-parameters (e.g. larger number of epochs, different learning rates and schedulers), but they do not produce better results.

\subsection{Evaluation}
\subsubsection{Multi-Instrument Transcription F1 Score}
Following MT3, we report the multi-instrument F1 score. More specifically, we ignore the offset and focus on the onset F1 score, as argued by \cite{lu2023multitrack} that not all musical instruments have a well-defined offset. As in MT3, we report the multi-instrument F1 scores in three levels of instrument granularity, namely ``Flat" (treat all non-drum instruments as a single instrument), ``MIDI Class" (maps
instruments to their MIDI class), and ``Full" (retains instruments’
program numbers as transcribed).  For simplicity, we denote multi-instrument transcription onset F1 score as \textbf{onset F1 scores} in the following discussions. For more details regarding the evaluation framework of multi-instrument transcription models, we refer the readers to section 4.2 of the MT3 paper \cite{gardner2021mt3}.

\begin{table*}[htbp]
  \centering
  \begin{tabular}{lccccccccc}
    \toprule
    & & & \multicolumn{3}{c}{Onset F1 Scores (\%)} & & \multicolumn{3}{c}{Instrument Detection (\%)}\\
    \cmidrule{4-6} \cmidrule{8-10}
    \text{Model} & $L_\text{agg}$ & $L_\text{max\_hop}$  & \text{Flat} & \text{MIDI Class} & \text{Full} & \text{$\phi$} & \text{Precision} & \text{Recall} & \text{F1} \\
    \hline
    MT3 (fixed) & N.A & N.A & 72.5 & 68.8 & 37.9 & 1.40 & 33.4 & \textbf{45.5} & 37.8\\
    MT3 (shuffled) & N.A & N.A & 72.1 & 68.6 & 38.1 & 1.30 & 35.4 & 45.2 & 39.1\\
    MR-MT3 (fixed) & 64 & 3 & \textbf{73.5} & \textbf{70.0} & 38.4 & 1.18 & 38.2 & 44.4 & 40.5\\
    MR-MT3 (shuffled) & 64 & 3 & 73.0 & 69.6 & \textbf{38.7} & \textbf{1.12} & \textbf{39.9} & 43.8 & \textbf{41.1} \\
    \bottomrule
  \end{tabular}
  \caption{Result for continuing the model from the checkpoint for 100 epochs}
  \label{tab:result_continual}
\end{table*}
\subsubsection{Instrument Leakage Ratio}
While higher F1 scores usually imply a better transcription result, they are not sufficient to inform about the overall transcription quality, e.g. a transcription with high F1 scores might have overly fragmented instrumentation.
To measure this phenomenon, we propose a new metric $\phi$, which measures the \textbf{instrument leakage ratio}. Given each sample $s_i$ in the test set $S$,  we denote the set of instruments in a transcribed MIDI as $I^{s_i}_\text{tr}$, and the set of instruments in the ground truth as $I^{s_i}_\text{gt}$. The instrument leakage ratio $\phi$ on the test set is defined as:

\begin{equation}\label{eq:leakage}
    \phi = \frac{\displaystyle\sum\limits_{i}^{|S|} |I^{s_i}_\text{tr}|}{\displaystyle\sum\limits_{i}^{|S|} |I^{s_i}_\text{gt}|}
\end{equation}
When the $\phi=1$, the model is able to detect the number of instruments correctly; if $\phi>1$, it implies that the model has the tendency to predict more instruments; if $\phi<1$, the model is underpredicting the number of instruments. 

\subsubsection{Instrument Detection F1 Score}
Although $\phi$ quantifies whether the models consistently over-predict or under-predict the number of instruments, it does not offer insights into the accuracy of the predicted instruments. To bridge this gap,  we introduce the \textbf{instrument detection F1 score}. Using the same previously defined $I_\text{tr}$ and $I_\text{gt}$ as above, we calculate precision, recall and F1 score as follows: 
\begin{equation}
    \text{Precision} = \frac{|I_\text{tr} \cap I_\text{gt}|}{|I_\text{tr}|}
\end{equation}

\begin{equation}
    \text{Recall} = \frac{|I_\text{tr} \cap I_\text{gt}|}{|I_\text{gt}|}
\end{equation}

\begin{equation}
    \text{F1} = \frac{2 \times \text{Precision} \times \text{Recall}}{\text{Precision} \times \text{Recall}}
\end{equation}

A model with a high $\phi$ tends to exhibit a low precision. In other words, mitigating the issue of instrument leakage necessitates an improvement in the precision and F1 score of instrument detection.

\begin{table*}[htbp]
  \centering
  \begin{tabular}{l|ccc|ccc|ccc}
    \toprule
    Models & \multicolumn{3}{c}{Slakh} & \multicolumn{3}{c}{ComMu} & \multicolumn{3}{c}{NSynth} \\\hline
    & Flat & MIDI Class  & Full & Flat & MIDI Class  & Full & Flat & MIDI Class  & Full \\\hline
    MT3 ComMu & 15.7 & 3.2 & 0.8 & \textbf{92.2} & \textbf{91.3} & \textbf{91.2} & 34.8 & 8.9 & 2.8\\
    MT3 Slakh Mixes ($\phi$ =1.65) & 64.9 & 59.5 & 33.5 & 51.3 & 35.3 & 32.1 & 19.4 & 7.7 & 3.8 \\ 
    MT3 Slakh Mixes + Stems ($\phi$ =1.43) & 59.6 & 53.8 & 30.5 & 59.0 & 43.2 & 30.5 & 34.6 & \textbf{14.7} & \textbf{6.4} \\ 
    MR-MT3 Slakh Mixes ($\phi$ =1.24) & \textbf{67.3} & \textbf{62.5} & \textbf{35.3} & 48.8 & 33.0 & 29.8 & 24.0 & 9.1 & 4.5 \\ 
    MR-MT3 Slakh Mixes + Stems {($\phi$ = 0.79)} & {62.3} & {57.2} & {32.2} & {53.1} & {37.2} & {20.6} & \textbf{{35.7}} & {13.5} & {5.2} \\

    \bottomrule
  \end{tabular}
  \caption{Cross-dataset evaluation. All models are trained from scratch using either ComMu or Slakh, and they are evaluated across Slakh, ComMu, and NSynth. The numbers in parenthesis next to the model names represent instrument leakage ratio.}
  \label{tab:domain_overfitting}
\end{table*}
\section{Results}\label{sec:results}
\subsection{From Scratch}\label{sec:results_scratch}
Our experimental results show that using token shuffling during training improves the onset F1 scores for all granularities (the first and second rows in Table~\ref{tab:result_scratch}).  
This evidence suggests that token shuffling is a beneficial data augmentation technique for improving the quality of token-based transcription, as opposed to using fixed-order tokens.

Upon the integration of the memory retention mechanism into MT3, in conjunction with token shuffling, we observe a further improvement in the F1 scores and a reduction in the instrument leakage phenomenon. When $L_\text{agg} = 32$ (the third row of Table~\ref{tab:result_scratch}), the MIDI Class F1 score is improved by 2.1 percentage points (pp). to 61.6\%, and the Full F1 score is improved by 1.0 pp to 34.5\%. Meanwhile, $\phi$ is significantly reduced from 1.65 to 1.05, and the instrument detection F1 score improves from 34.2\% to 39.5\%. When $L_\text{agg}$ is increased to 64 (the fourth row of Table~\ref{tab:result_scratch}), there are improvements of around 0.5\% on all F1 scores, while keeping a relatively good $\phi$ and instrument detection F1 scores. Due to the diminishing returns when further increasing $L_\text{agg}$, we choose 64 as the best parameter and continue exploring the sampling of prior segments. These findings highlight the effectiveness of the memory retention mechanism. Leveraging past event tokens is able to better detect the instruments appear in the current segment. This approach improves the accuracy of instrument detection, and helps to reduce instrument leakage.

Apart from using the immediate prior frames as memory ($L_\text{max\_hop} = 1$), we also experiment with sampling prior frames from a more distant past. When $L_\text{max\_hop}$ is increased to 3, the F1 scores show further improvement. However, when $L_\text{max\_hop}$ is increased up to 8, the F1 scores cease to improve. We attribute this to the decrease in musical relevance of the past segments in relation to the current segment, as the inclusion of irrelevant information does not help the model with transcribing the current segment. We also observe that further increasing $L_\text{max\_hop}$ begins to introduce more instrument leakage and negatively impacts instrument detection F1 scores.
Nonetheless, the results in Table~\ref{tab:result_scratch} indicate that our proposed methods are effective at improving both the onset F1 scores and the instrument detection F1 score, while keeping $\phi$ at a lower value than the baseline.  

\subsection{Continual Training}\label{sec:results_continual}
Transitioning from training the models from scratch, we apply continual training to the model using the official checkpoint as the initial weight. We observe a similar trend in this phase as well. Comparing the results of fixed token ordering (first and third rows) and token shuffling (second and fourth rows) in Table~\ref{tab:result_continual}, we observe that token shuffling is effective in improving the instrument detection F1 score and mitigating instrument leakage. 

Upon applying our proposed method with the optimal parameters ($L_\text{agg}=64$, $L_\text{max\_hop}=3$) derived from Table~\ref{tab:result_scratch}, we achieve an improvement of around 1 pp in onset F1 scores and instrument detection F1, and also significantly reduce $\phi$. This aligns with the observations made in Section~\ref{sec:results_scratch}. Interestingly, we notice a subtle trade-off between transcription accuracy and the instrument detection F1 score. From the third and fourth rows of~Table~\ref{tab:result_continual}, we can see that higher onset F1 scores do not necessarily result in lower instrument leakage. We also observe the same trend in Table~\ref{tab:result_scratch}, whereby $\phi$ might increase together with the onset F1 scores. We believe that to further advance the field of multi-instrument music transcription, it is essential to shift focus from merely improving transcription F1 scores to enhancing the instrument detection F1 score. Future research should aim to develop a method that can concurrently improve these metrics without compromising on any of them. Regardless, these results indicate that our proposed methods are effective in mitigating the instrument leakage issue.

\subsection{Domain Overfitting}
Given the relative complexity of multi-instrument music transcription in comparison to pitch detection and monophonic music transcription, it is anticipated that a proficient multi-instrument transcription model would demonstrate superior performance in the single-instrument, monophonic audio setting. Contrary to expectations, models trained on datasets with a higher degree of complexity, such as Slakh, do not necessarily perform well on simpler datasets. To validate this observation, experiments were conducted using two single-instrument datasets: ComMU and NSynth.
\begin{itemize}
  \item ComMU \cite{lee2022commu}: ComMU provides MIDI stems composed for dedicated instruments. We synthesize the corresponding audio using \texttt{pretty\_midi.fluidsynth}, based on the annotated instrument program number. We further split the dataset with a train/val/test ratio of 90:5:5.
  \item NSynth \cite{nsynth2017}: NSynth provides audio tracks of single musical notes across 10 instrument families (excluding vocals). We generate the corresponding MIDI based on the annotated note number, and use the train/val/test splits in the dataset.
\end{itemize}

The results, as illustrated in Table~\ref{tab:domain_overfitting}, indicate that a model trained on a simpler dataset like ComMU performs suboptimally on Slakh, a dataset with a higher degree of complexity. This aligns with expectations. Interestingly, when the model is trained on Slakh audio mixtures, its performance on ComMU is not satisfactory, with F1 scores even lower than those for Slakh. This suggests the model struggles to determine which instrument is playing the notes. A similar observation can be made with the NSynth dataset.

It is suspected that MT3 overfits to audio mixtures. Therefore, an attempt was made to train MT3 with stems. However, this resulted in a lower onset F1 score for the Slakh test set. When evaluated on ComMU and NSynth, the models trained with both audio mixes and stems have significantly higher flat and MIDI class onset F1 scores. Along with a lower instrument leakage achieved, this shows that training with audio stems aids in preventing overfitting.

Upon applying the proposed method on top of the audio mixes and stems training, it was observed that the instrument leakage ratio is only 0.79, implying that the model is predicting fewer instruments than there should be. As before, the onset F1 scores for the Slakh dataset are lower, while the F1 scores for ComMU and NSynth are higher. However, it appears that the model may be over-regularized by applying two techniques simultaneously, as the F1 scores for ComMU and NSynth obtained with the proposed ideas are not as high as the baseline. This aspect is left for future work.

\section{Conclusion and Future Work}

In this study, we propose MR-MT3, an enhanced version of the token based multi-instrument automatic music transcription (AMT) model, MT3, with the aim of addressing existing issues. The introduced \textbf{memory retention mechanism} leverages past musical events to capture long-term context, demonstrating its effectiveness in mitigating issues such as instrument leakage. Our \textbf{prior token sampling} provides a flexible approach to inform the transcription of current frames by sampling relevant prior tokens up to a certain time period in the past. Additionally, \textbf{token shuffling} serves as an effective data augmentation technique, contributing to improved transcription results. Through comprehensive experiments involving training from scratch and continual training from the MT3 checkpoint, we demonstrate the efficacy of our proposed methods in enhancing multi-instrument transcription onset F1 scores, instrument detection F1 scores, and mitigating instrument leakage. We hope to encourage further research into multi-instrument AMT that considers overall transcription quality beyond the note-level F1 score.

Moreover, our investigation into domain overfitting uncovers difficulties adapting models trained on complex multi-instrument datasets to simpler single-instrument datasets. This highlights the importance of improving model generalization across various transcription scenarios. Suitable data augmentation or domain adaption techniques might help to alleviate this issue. Future research directions may include further exploration of domain adaptation strategies and the integration of additional evaluation metrics to comprehensively assess multi-instrument transcription models.






\bibliographystyle{named}
\bibliography{ijcai24}

\begin{thebibliography}{}

\bibitem[\protect\citeauthoryear{Benetos \bgroup \em et al.\egroup }{2019}]{benetos2018automatic}
Emmanouil Benetos, Simon Dixon, Zhiyao Duan, and Sebastian Ewert.
\newblock Automatic music transcription: An overview.
\newblock {\em IEEE Signal Processing Magazine}, 36:20--30, 2019.

\bibitem[\protect\citeauthoryear{Bittner \bgroup \em et al.\egroup }{2017}]{bittner2017deep}
Rachel~M. Bittner, Brian McFee, Justin Salamon, Peter Li, and Juan~P. Bello.
\newblock Deep salience representations for {F0} estimation in polyphonic music.
\newblock In {\em ISMIR}, 2017.

\bibitem[\protect\citeauthoryear{Bittner \bgroup \em et al.\egroup }{2022}]{bittner2022lightweight}
Rachel~M. Bittner, Juan~Jos{\'e} Bosch, David Rubinstein, Gabriel Meseguer-Brocal, and Sebastian Ewert.
\newblock A lightweight instrument-agnostic model for polyphonic note transcription and multipitch estimation.
\newblock In {\em ICASSP}, 2022.

\bibitem[\protect\citeauthoryear{Chen \bgroup \em et al.\egroup }{2021}]{chen2021pix2seq}
Ting Chen, Saurabh Saxena, Lala Li, David~J Fleet, and Geoffrey Hinton.
\newblock Pix2seq: A language modeling framework for object detection.
\newblock In {\em ICLR}, 2021.

\bibitem[\protect\citeauthoryear{Cheuk \bgroup \em et al.\egroup }{2021}]{cheuk2021reconvat}
Kin~Wai Cheuk, Dorien Herremans, and Li~Su.
\newblock {ReconVAT}: A semi-supervised automatic music transcription framework for low-resource real-world data.
\newblock In {\em ACM MM}, 2021.

\bibitem[\protect\citeauthoryear{Cheuk \bgroup \em et al.\egroup }{2023a}]{cheuk2023jointist}
Kin~Wai Cheuk, Keunwoo Choi, Qiuqiang Kong, Bochen Li, Minz Won, Ju-Chiang Wang, and Yun-Ning Hung~Dorien Herremans.
\newblock Jointist: Simultaneous improvement of multi-instrument transcription and music source separation via joint training.
\newblock {\em arXiv preprint arXiv:2302.00286}, 2023.

\bibitem[\protect\citeauthoryear{Cheuk \bgroup \em et al.\egroup }{2023b}]{cheuk2023diffroll}
Kin~Wai Cheuk, Ryosuke Sawata, Toshimitsu Uesaka, Naoki Murata, Naoya Takahashi, Shusuke Takahashi, Dorien Herremans, and Yuki Mitsufuji.
\newblock Diffroll: Diffusion-based generative music transcription with unsupervised pretraining capability.
\newblock In {\em ICASSP 2023-2023 IEEE International Conference on Acoustics, Speech and Signal Processing (ICASSP)}, pages 1--5. IEEE, 2023.

\bibitem[\protect\citeauthoryear{Cwitkowitz \bgroup \em et al.\egroup }{2024}]{cwitkowitz2023timbre}
Frank Cwitkowitz, Kin~Wai Cheuk, Woosung Choi, Marco~A Mart{\'\i}nez-Ram{\'\i}rez, Keisuke Toyama, Wei-Hsiang Liao, and Yuki Mitsufuji.
\newblock Timbre-trap: A low-resource framework for instrument-agnostic music transcription.
\newblock In {\em ICASSP 2024-2024 IEEE International Conference on Acoustics, Speech and Signal Processing (ICASSP)}, page in press. IEEE, 2024.

\bibitem[\protect\citeauthoryear{Engel \bgroup \em et al.\egroup }{2017}]{nsynth2017}
Jesse Engel, Cinjon Resnick, Adam Roberts, Sander Dieleman, Douglas Eck, Karen Simonyan, and Mohammad Norouzi.
\newblock Neural audio synthesis of musical notes with wavenet autoencoders, 2017.

\bibitem[\protect\citeauthoryear{Ewert \bgroup \em et al.\egroup }{2012}]{Ewert}
Sebastian Ewert, Meinard Muller, Verena Konz, Daniel Mullensiefen, and Geraint~A. Wiggins.
\newblock Towards cross-version harmonic analysis of music.
\newblock {\em IEEE Transactions on Multimedia}, 14(3):770--782, 2012.

\bibitem[\protect\citeauthoryear{Gardner \bgroup \em et al.\egroup }{2022}]{gardner2021mt3}
Josh Gardner, Ian Simon, Ethan Manilow, Curtis Hawthorne, and Jesse Engel.
\newblock {MT3}: Multi-task multitrack music transcription.
\newblock In {\em ICLR}, 2022.

\bibitem[\protect\citeauthoryear{Hawthorne \bgroup \em et al.\egroup }{2018}]{hawthorne2017onsets}
Curtis Hawthorne, Erich Elsen, Jialin Song, Adam Roberts, Ian Simon, Colin Raffel, Jesse Engel, Sageev Oore, and Douglas Eck.
\newblock Onsets and frames: Dual-objective piano transcription.
\newblock In {\em ISMIR}, 2018.

\bibitem[\protect\citeauthoryear{Hawthorne \bgroup \em et al.\egroup }{2021}]{Hawthorne2021SequencetoSequencePT}
Curtis Hawthorne, Ian Simon, Rigel Swavely, Ethan Manilow, and Jesse Engel.
\newblock Sequence-to-sequence piano transcription with transformers.
\newblock In {\em ISMIR}, 2021.

\bibitem[\protect\citeauthoryear{Hawthorne \bgroup \em et al.\egroup }{2022}]{hawthorne2022multi}
Curtis Hawthorne, Ian Simon, Adam Roberts, Neil Zeghidour, Josh Gardner, Ethan Manilow, and Jesse Engel.
\newblock Multi-instrument music synthesis with spectrogram diffusion.
\newblock {\em ISMIR}, 2022.

\bibitem[\protect\citeauthoryear{Kelz \bgroup \em et al.\egroup }{2019}]{kelz2019deep}
Rainer Kelz, Sebastian B{\"o}ck, and Gerhard Widmer.
\newblock Deep polyphonic adsr piano note transcription.
\newblock In {\em ICASSP 2019-2019 IEEE International Conference on Acoustics, Speech and Signal Processing (ICASSP)}, pages 246--250. IEEE, 2019.

\bibitem[\protect\citeauthoryear{Kong \bgroup \em et al.\egroup }{2021}]{kong2021high}
Qiuqiang Kong, Bochen Li, Xuchen Song, Yuan Wan, and Yuxuan Wang.
\newblock High-resolution piano transcription with pedals by regressing onset and offset times.
\newblock {\em IEEE/ACM Transactions on Audio, Speech, and Language Processing}, 29:3707--3717, 2021.

\bibitem[\protect\citeauthoryear{Lee \bgroup \em et al.\egroup }{2022}]{lee2022commu}
Hyun Lee, Taehyun Kim, Hyolim Kang, Minjoo Ki, Hyeonchan Hwang, Sharang Han, Seon~Joo Kim, et~al.
\newblock Commu: Dataset for combinatorial music generation.
\newblock {\em Advances in Neural Information Processing Systems}, 35:39103--39114, 2022.

\bibitem[\protect\citeauthoryear{Lei \bgroup \em et al.\egroup }{2020}]{lei2020mart}
Jie Lei, Liwei Wang, Yelong Shen, Dong Yu, Tamara~L Berg, and Mohit Bansal.
\newblock Mart: Memory-augmented recurrent transformer for coherent video paragraph captioning.
\newblock {\em arXiv preprint arXiv:2005.05402}, 2020.

\bibitem[\protect\citeauthoryear{Lu \bgroup \em et al.\egroup }{2023}]{lu2023multitrack}
Wei-Tsung Lu, Ju-Chiang Wang, and Yun-Ning Hung.
\newblock Multitrack music transcription with a time-frequency perceiver.
\newblock In {\em ICASSP}, 2023.

\bibitem[\protect\citeauthoryear{Pedersoli \bgroup \em et al.\egroup }{2020}]{pedersoli2020improving}
Fabrizio Pedersoli, George Tzanetakis, and Kwang~Moo Yi.
\newblock Improving music transcription by pre-stacking a {U-Net}.
\newblock In {\em ICASSP}, 2020.

\bibitem[\protect\citeauthoryear{Ru}{2021}]{ru2021computer}
Yingdong Ru.
\newblock Computer assisted chord detection using deep learning and yolov4 neural network model.
\newblock In {\em Journal of Physics: Conference Series}, volume 2083, page 042017. IOP Publishing, 2021.

\bibitem[\protect\citeauthoryear{Tanaka \bgroup \em et al.\egroup }{2020}]{tanaka2020multi}
Keitaro Tanaka, Takayuki Nakatsuka, Ryo Nishikimi, Kazuyoshi Yoshii, and Shigeo Morishima.
\newblock Multi-instrument music transcription based on deep spherical clustering of spectrograms and pitchgrams.
\newblock In {\em ISMIR}, pages 327--334, 2020.

\bibitem[\protect\citeauthoryear{Thickstun \bgroup \em et al.\egroup }{2017}]{thickstun2016learning}
John Thickstun, Zaid Harchaoui, and Sham Kakade.
\newblock Learning features of music from scratch.
\newblock In {\em ICLR}, 2017.

\bibitem[\protect\citeauthoryear{Wongsaroj \bgroup \em et al.\egroup }{2014}]{Wongsaroj}
Chaisup Wongsaroj, Nakornthip Prompoon, and Athasit Surarerks.
\newblock A music similarity measure based on chord progression and song segmentation analysis.
\newblock In {\em 2014 Fourth International Conference on Digital Information and Communication Technology and its Applications (DICTAP)}, pages 158--163, 2014.

\bibitem[\protect\citeauthoryear{Wu \bgroup \em et al.\egroup }{2019}]{wu2019polyphonic}
Yu-Te Wu, Berlin Chen, and Li~Su.
\newblock Polyphonic music transcription with semantic segmentation.
\newblock In {\em ICASSP 2019-2019 IEEE International Conference on Acoustics, Speech and Signal Processing (ICASSP)}, pages 166--170. IEEE, 2019.

\bibitem[\protect\citeauthoryear{Wu \bgroup \em et al.\egroup }{2020}]{wu2020multi}
Yu-Te Wu, Berlin Chen, and Li~Su.
\newblock Multi-instrument automatic music transcription with self-attention-based instance segmentation.
\newblock {\em IEEE/ACM Transactions on Audio, Speech, and Language Processing (TASLP)}, 28:2796--2809, 2020.

\bibitem[\protect\citeauthoryear{Wu \bgroup \em et al.\egroup }{2022}]{wu2022memvit}
Chao-Yuan Wu, Yanghao Li, Karttikeya Mangalam, Haoqi Fan, Bo~Xiong, Jitendra Malik, and Christoph Feichtenhofer.
\newblock Memvit: Memory-augmented multiscale vision transformer for efficient long-term video recognition.
\newblock In {\em Proceedings of the IEEE/CVF Conference on Computer Vision and Pattern Recognition}, pages 13587--13597, 2022.

\end{thebibliography}

\end{document}